# Anomalous $k$-dependent spin splitting in wurtzite $Al_xGa_{1-x}N$/GaN heterostructures

**Ikai Lo,[a)] M. H. Gau, J. K. Tsai, Y. L. Chen, Z. J. Chang, W. T. Wang, and J. C. Chiang**

Department of Physics, Center for Nanoscience and Nanotechnology, National Sun Yat-Sen University, Kaohsiung, Taiwan, Republic of China

**T. Aggerstam**

KTH Royal Institute of Technology, Electrum 229, SE-164 40 Kista, Sweden

Abstract

We have confirmed the $k$-dependent spin splitting in wurtzite $Al_xGa_{1-x}N$/GaN heterostructures. Anomalous beating pattern in Shubnikov-de Haas measurements arises from the interference of Rashba and Dresselhaus spin-orbit interactions. The dominant mechanism for the $k$-dependent spin splitting at high values of $k$ is attributed to Dresselhaus term which is enhanced by the $\Delta_{C1}$-$\Delta_{C3}$ coupling of wurtzite band folding effect.

*Introduction (interference of Rashba and Dresselhaus effects).* Spin-orbit interaction is the key issue of spin dynamics in semiconductor spintronics [1,2]. It originates from the relativistic theory of Dirac's equation to assure symmetry in the wave equation with respect to space and time derivatives [3]. In the atomic case, it is formulated as $H_{SO}=(ge\hbar/4m_0^2c^2)\vec{\sigma}\cdot(\vec{p}\times\nabla V)$ for the electron spin interacting with the electric field ($E=-\nabla V$) produced by the nucleus with respect to its orbital motion, where $\vec{\sigma}$ is Pauli matrix and $\vec{p}$ the momentum of electron. In III-V semiconductor heterostructures, there are two effects contributing to the spin-orbit interaction: (i) the structure inversion asymmetry (SIA) of electrostatic confinement potential at hetero-interface, i.e., $\nabla V = \hat{z}(dV/dz)$, (known as Rashba effect), and (ii) the bulk inversion asymmetry (BIA) of the lattice structure in which the coupling between *s*-wave of atom A and *p*-wave of atom B is not equal to that between *p*-wave of atom A and *s*-wave of atom B, i.e., U($s_A,p_B$)-U($p_A,s_B$)≠0, (named Dresselhaus effect). In ***zincblende*** structure, the 2D Hamiltonian for SIA term can be written as $H_R(k) = \alpha(\sigma_x k_y - \sigma_y k_x)$, where Rashba coefficient $\alpha$ is proportional to the electric field built at hetero-interface; $\alpha = \alpha_0 eE$, [4]. While, for BIA term it becomes $H_D(k) = \gamma[<k_z^2>(\sigma_y k_y - \sigma_x k_x) + (\sigma_x k_x k_y^2 - \sigma_y k_y k_x^2)]$, where *z*-axis is the growth direction and Dresselhaus constant ($\gamma$) is proportional to the strength of BIA spin-orbit interaction [5-8]. If only consider the linear-*k* terms, then $H_D$ and $H_R$ are coupled by each other, leading to a linear-***k***-dependent interference: $H_{SO}(k) = H_R(k) + H_D(k) = \alpha(\sigma_x k_y - \sigma_y k_x) + \beta(\sigma_x k_x - \sigma_y k_y)$, where $\beta = -\gamma<k_z^2> = -\gamma(\pi/d)^2$ and *d* is the size of quantum confinement [9,10]. The degree of specular asymmetry of two-dimensional electron gas (2DEG) at hetero-interface can be controlled by a gate voltage, and thus the relative contributions of the two effects are tunable [10-12]. The spin splitting becomes anisotropic when the two contributions are comparable, giving an anomalous beating pattern in magneto-oscillations, e.g., Shubnikov-de Haas (SdH) oscillations. The suppression of beating pattern due to the ***Interference of Rashba and Dresselhaus effects*** was individually proposed by Tarasenko et al. [11] and by Ting et al. [10]. Interplay between Rashba and Dresselhaus effects activates a crucial impact in spin dynamics [9], and a large Dresselhaus term provides significant enhancement for polarization efficiency in spintronic application [10]. Recently, the candidate material for large Dresselhaus term is focused on III-nitride compounds due to their inherent property of bulk structure inversion asymmetry in z-direction [13].

*Spin-orbit interaction in **wurtzite GaN**.*   The effective spin-splitting energy $\Delta E^*_{SO}(k)$ measured by SdH oscillations is widely spread from 0 to 11 meV [13-17] in wurtize $Al_xGa_{1-x}N/GaN$, Table I.  In theory, the calculated Rashba spin splitting by four-band model ($\Delta E_R < 1$ meV) is much smaller than the experimental values [18].  On the other hand, the Dresselhaus effect was evaluated in wurtzite lattice with intrinsic bulk inversion asymmetry by Lew Yan Voon et al. [19]. It was pointed out that the linear-$k$ term can also arise from the weak $s$-$p_z$ mixing of conduction band at $k = 0$ due to the second Ga-N neighbor interaction [20].  However, the Dresselhaus spin-splitting energy calculated by four-band model is very small (e.g., $\Delta E_D < 0.5$ meV and 0.05 meV for CdS and ZnO, respectively) [19].  Therefore, the Rashba and Dresselhaus effects, which consider the spin-orbit interaction between conduction and valence bands (the four-band model), are not enough to account for the large spin splitting observed in $Al_xGa_{1-x}N/GaN$. Recently, we proposed a model of $\Delta_{C1}$-$\Delta_{C3}$ coupling to interpret the large spin splitting, which is caused by a band folding effect due to the different size of unit cells in the hexagonal direction between zinc-blende and wurtzite lattices [21].  Therefore, we can extend the Rashba term to include the intrinsic wurtzite structure inversion asymmetry (WSIA) effect [4,20], which leads to a linear-$k$ term, $\boldsymbol{H}_R(\boldsymbol{k}) = \alpha_{wz}(\sigma_x k_y - \sigma_y k_x)$ with $\alpha_{wz} = \alpha_0 eE + C_{WSIA}$.  Here the Rashba coefficient ($\alpha_{wz}$) consists of the WSIA constant ($C_{WSIA}$) as well as the electric field built at hetero-interface ($\boldsymbol{E}$).  In the meanwhile, the Dresselhaus term becomes $\boldsymbol{H}_D(\boldsymbol{k}) = \gamma_{wz}(k_{//}^2 - bk_z^2)(\sigma_x k_y - \sigma_y k_x)$, where $b = 4$ for ideal wurtzite.  Both $\boldsymbol{H}_R$ and $\boldsymbol{H}_D$ are proportional to $(\sigma_x k_y - \sigma_y k_x)$.  This isotropic behavior was recently demonstrated to evaluate the Rashba term in GaN/AlGaN [22].  However, the Rashba term itself has only a spin-degeneracy at Γ-point ($\boldsymbol{k} = 0$), implying that the observation of spin-degenerate states other than Γ-point evidences the presence of Dresselhaus term.  The Hamiltonian for the $\boldsymbol{H}_R$ and $\boldsymbol{H}_D$ interference becomes: $\boldsymbol{H}_{SO}(\boldsymbol{k}) = [\alpha_{wz} - \gamma_{wz}(bk_z^2 - k_{//}^2)](\sigma_x k_y - \sigma_y k_x)$.  The suppression of beating pattern due to the interference of $\boldsymbol{H}_D$ and $\boldsymbol{H}_R$ turns into a spin-degenerate sandglass cone with a degenerate equation: $[\alpha_{wz} - \gamma_{wz}(bk_z^2 - k_{//}^2)] = 0$.  The schematic spin-degenerate sandglass surface for a bulk wurtizte structure with constant $\alpha_{wz}$ and $\gamma_{wz}$ is plotted in Fig. 1 (a). Besides, in $Al_xGa_{1-x}N/GaN$ the piezoelectric field built at hetero-interface [23] can enhance the Rashba effect, as well. Theerefore, the mechanism of spin splitting consists of four parts: Rashba ($\alpha_{wz}$) term, Dresselhaus ($\beta_{wz}$) term, piezoelectric field and $\Delta_{C1}$-$\Delta_{C3}$ coupling.

*Experiments.* The $Al_xGa_{1-x}N/GaN$ samples were grown by metal-organic chemical vapor deposition on a *c*-plane sapphire with ferrocene as an iron doping source. Above the sapphire substrate, the sample consists of a 100-nm-thick GaN buffer layer, an 800-nm-thick Fe-doped GaN layer, an unintentionally doped 1.6-μm-thick GaN layer, and an unintentionally doped 25-nm-thick $Al_xGa_{1-x}N$ barrier layer on the top. The Fe-doped impurities are used to form deep-level electron traps to capture the *n*-type carriers as an insulating buffer (the activation energy is 0.5 eV) [24]. A Hall-bar-shaped 2 mm x 6 mm sample ($x$ = 0.22) with indium ohmic contacts was annealed at 350 °C for 5 minutes under $N_2$ forming gas. The persistent photoconductivity (PPC) effect was provided by illuminating the sample at T ~ 0.38 K for different time periods using a blue light-emitting diode of 472-nm wavelength. Before illumination, the carrier concentration is 9.1 x $10^{12}$ cm$^{-2}$ and the mobility is 8.1 x $10^3$ cm$^2$/Vs (determined from Hall measurement at T ~ 0.38 K). After an extensive illumination, the carrier concentration increases by 14% to 10.4 x $10^{12}$ cm$^{-2}$, but the mobility decreases to 6.3 x $10^3$ cm$^2$/Vs. The increased carriers were mostly transferred from the deep-level electron traps located at the remote Fe-doped GaN layer [25]. We performed the SdH measurement on the sample for the magnetic field ($B$) from 0.5 to 12 T. Because the oscillation part of magneto-resistance ($R_{XX}$) is a cosine function against $1/B$, the SdH data (2048 points) were taken with equal spacing of $1/B$ for the purpose of fast Fourier transformation (FFT). The resolution of SdH frequency in the FFT spectrum is equal to 0.26 T, which can detect the change of 0.13 x $10^{11}$cm$^{-2}$ in 2D carrier concentration. The SdH measurements and their FFT spectra at different temperatures before illumination are shown in Fig. 1(b) and (d). A single oscillation is detected at frequency $f_1$ = 179.27 T, giving the carrier concentration of first subband $n_1 = 2f_1e/h$ = 8.68 x $10^{12}$ cm$^{-2}$, and Zeeman splitting is observed at high fields (e.g., $B$ > 10 T) for temperatures lower than 2.20 K [26]. Since the first subband carrier concentration is less than the Hall carrier concentration, the excess carriers may reside in the second subband. Because of the greater mobility, only the SdH oscillation of first subband is detected. It is confirmed by the measured effective mass, extracted from the temperature dependence of SdH amplitudes at the fields of peak positions, Fig. 1(c). The average value of the mass ($m^*/m_0 = 0.215 \pm 0.003$) is equal to the first subband effective mass [15].

*Beating pattern of SdH oscillations.* Figure 2 shows the SdH measurements and their FFT spectra after the PPC effect for different illumination times. The spin splitting of the first subband (i.e., $f_\uparrow$ and $f_\downarrow$) appears after 71s illumination, and the peaks separate farther for a longer illumination until saturation, Fig. 2(b). The amplitude of spin-down oscillation ($f_\downarrow$) decreases with the illumination time, but that of spin-up oscillation ($f_\uparrow$) increases, relatively. Zeeman splitting also appears at high fields for 3s and 71s illuminations. Zeeman splitting has been observed by Cho et al. [26] and analyzed in the presence of Rashba and Dresselhaus effects by de Andrada e Silva et al. [27]. It was shown that the Zeeman splitting can be detected at high fields as the Zeeman energy is greater than the energy of spin-orbit interaction. The effective $g$-factor is about 2 for GaN [26], and hence the Zeeman splitting for $B > 10$ T ($\Delta E_Z > 0.58$ meV) is greater than the effective spin-splitting energy ($\Delta E^*_{SO}$) in the first two measurements (3s and 71s). Since Zeeman splitting is linearly dependent on $B$ ($\Delta E_Z = \frac{1}{2} g^* \mu_B B$), it produces only a phase constant in SdH oscillation and will not change the frequency for large Landau levels like our case. As a result, Zeeman splitting enhances the amplitude of second harmonic peak in the FFT spectrum (see the peak at $2f_1$ frequency in Fig. 1b). In Fig. 2(b), a third peak ($f_{MIS}$) is growing between the two spin-splitting peaks in the longer illuminations. A three-peak SdH pattern was reported by Tsubaki et al. [14], and by Averkiev et al. [28]. The former attributed the third peak to a magneto-intersubband scattering (MIS) component, and the latter ascribed it to the third harmonics of the interference of Rashba and Dresselhaus terms. The MIS component is a second-order resonant scattering oscillation between the first ($f_1$) and second ($f_2$) subbands with a frequency $f_{MIS} = f_1 - f_2$, and its amplitude does not contain the temperature damping factor.

Without the temperature damping factor, the MIS component is less sensitive to the change of temperature. We checked the SdH oscillations at different temperatures after 12671s illumination and showed in Figs. 2(c) and (d). It is evident that there is no change in the frequencies when the temperature increases. The amplitude of $f_{MIS}$ peak has less sensitivity on temperature, but the temperature-sensitivity for $f_\uparrow$ and $f_\downarrow$ peaks is consistent with that of $f_1$ peak in Fig. 1(d). Besides, the value of $f_{MIS}$ frequency is approximately equal to the difference of $f_\uparrow$ and $f_2$ frequencies. Therefore we conclude that the peak of $f_{MIS}$ is due to the MIS component between the $f_\uparrow$ and $f_2$ oscillations. In order to evaluate these peaks more precisely against the illumination time, we repeated the SdH measurements after warming up the sample at room temperature for a couple of days. Figure 3 shows the precise measurements at T ~ 0.38 K and their FFT spectra. The single SdH oscillation ($f_1$ peak) is detected for the illumination time less than 193 seconds, and it splits apparently after 313s illumination, Fig. 3(b). The two spin-splitting peaks separate farther by longer illuminations. The amplitude of $f_\downarrow$ peak decreases, but that of $f_\uparrow$ peak increases with the illumination time, as observed in Fig. 2. In the last two spectra it is clearly shown that the $f_{MIS}$ peak is split away from the $f_\uparrow$ peak. We calculated the carrier concentrations for the two spin subbands from the SdH frequencies and plotted them against the first subband carrier concentration (i.e., $n_1 = n_\uparrow + n_\downarrow$) in Fig. 4(a). The effective spin-splitting energy was calculated from the carrier concentrations of spin subbands, $\Delta E^*_{SO} = 2\pi\hbar^2(n_\uparrow - n_\downarrow)/m^*$, and plotted against the mean Fermi wave vector, $k_{//}^F = (2\pi n_1)^{1/2}$, as well. We obtained the $k$-dependent $\Delta E^*_{SO}$, and extrapolated it to zero at $k_{//}^F = 7.37 \times 10^8$ m$^{-1}$.

*Anomalous k-dependent spin-splitting energy* ($\Delta E^*_{SO}$). The 2D Hamiltonian of Rashba and Dresselhaus terms can be written as: $\boldsymbol{H}_{SO}(\boldsymbol{k}) = [\alpha_{wz} - \gamma_{wz}(b\langle k_z^2\rangle - k_{//}^2)](\sigma_x k_y - \sigma_y k_x)$, and the effective spin-splitting energy becomes $\Delta E^*_{SO}(k) = 2[(\alpha_{wz} - \gamma_{wz}b\langle k_z^2\rangle)k_{//} + \gamma_{wz}k_{//}^3]$. It can be separated into linear-$k_{//}$ and cubic-$k_{//}$ terms. If $\boldsymbol{H}_{SO}$ is solely governed by Rashba term ($\gamma_{wz} = 0$), a beating pattern of SdH oscillations arises from two isotropic spin-splitting concentric Fermi circles. However, if $[\alpha_{wz} - \gamma_{wz}(b\langle k_z^2\rangle - k_{//}^2)] = 0$, the two-spin subbands are degenerate (i.e., $\Delta E^*_{SO} = 0$), and hence the beating pattern does not appear (e.g., two-fold spin-degenerate with a constant $\langle k_z \rangle$ in Fig. 1a). As we mentioned before, the mechanism of spin splitting in $Al_xGa_{1-x}N/GaN$ consists of four parts: Rashba ($\alpha_{wz}$) term, Dresselhaus ($\gamma_{wz}$) term, piezoelectric field and $\Delta_{C1}$-$\Delta_{C3}$ coupling. Garrido et al. showed that the electric field $\boldsymbol{E}$ due to the piezoelectric and spontaneous polarization depends linearly on Al composition [23]. The Rashba coefficient $\alpha_{wz}$ thus increases linearly with Al composition, $\alpha_{wz}(x) = \alpha_0 e\boldsymbol{E} + \boldsymbol{C_{WSIA}}$. In the meanwhile, the stronger electric field gives a better quantum confinement (i.e., a smaller confinement size $d$). As a result, $\langle k_z^2 \rangle$ increases with Al composition as well, and hence the Fermi wave vector $\boldsymbol{k}^F = (k_{//}^F, k_z^F)$ of 2DEG will be pushed toward the anticrossing zone of $\Delta_{C1}$ and $\Delta_{C3}$ bands [21]. The spin-splitting energy (mainly Dresselhaus term) is tremendously enhanced by the $\Delta_{C1}$-$\Delta_{C3}$ coupling due to the increasd $p$-wave probability in the $\Delta_{C1}$ band, leading to an enhancement of effective Dresselhaus constant ($\gamma^*_{wz}$). As long as the spin-degenerate equation is held, the $\Delta E^*_{SO}(k)$ vanishes due to the interference of Rashba and Dresselhaus terms (i.e., no beating SdH oscillations). If the equation is broken with a variation of $k_{//}^F$ (e.g., by PPC effect here or by a gate voltage in [14]), the beating pattern occurs (e.g., observed in Fig. 2a and repeated in Fig. 3a). Since both $k_{//}^F$ and $\langle k_z^F \rangle$ are functions of 2D carrier concentration, we define a characteristic wave vector: the threshold wave vector ($k_{//}^{th}$), at which the two-fold degeneracy is just broken, leading to a non-vanishing $\Delta E^*_{SO}(k)$. In Fig. 4(a), we fit the $\Delta E^*_{SO}(k)$ versus $(k_{//}^F)^3$, the dotted line, and obtained the threshold wave vector $k_{//}^{th} = 7.37 \times 10^8$ m$^{-1}$. We also fit the $\Delta E^*_{SO}(k)$ versus linear-$k_{//}^F$ for comparison (the solid line). It is found that these two fits do not have much difference in the measured $k_{//}^F$. The threshold $k_{//}^{th}$ should be smaller than the value of $\langle k_z^F \rangle$. The size of quantum confinement can be written as $d = 2\langle z \rangle$, where $\langle z \rangle$ is the average distance of electron from the hetero-interface. We estimated the electric field $\boldsymbol{E}$ and the average distance $\langle z \rangle$ as functions of total sheet carrier concentration $n_T$ by a self-consistent solution of Poisson and Schrodinger equations. The material parameters for the piezoelectric coefficients in ref. [22] were used. The results of calculated $\langle z \rangle$ and $\boldsymbol{E}$ are shown in Fig. 4(b). The $\langle z \rangle$ is about 1.40 nm at $n_T = 8.65 \times 10^{12}$ cm$^{-2}$ and reduces to 1.29 nm at $n_T = 9.87 \times 10^{12}$ cm$^{-2}$, but the $\boldsymbol{E}$ increases from 1.22 to 1.36 MV/cm, respectively. Thus the calculated $\langle k_z^F \rangle$ is 1.12 $\times$ 10$^9$ m$^{-1}$, which is greater than the threshold wave vector as we predicted. The value of $\langle (k_z^F)^2 \rangle$ increases by 18%, but the $E$ increases only by 11%. Therefore, the degeneracy of two spin subbsnds was broken, and hence the Dresselhaus term starts to dominate $\Delta E^*_{SO}(k)$. The non-vanishing $\Delta E^*_{SO}(k)$ for beating pattern depends on the competition between Rashba ($\alpha_{wz}$) and Dresselhaus ($\beta_{wz}$) terms, resulting in the diverse $\Delta E^*_{SO}(k)$ observed in $Al_xGa_{1-x}N/GaN$ (Table I). It is noted that the Si-doping level at $Al_xGa_{1-x}N$ will alter both $E$ and $<z>$, yielding a fluctuation in $\Delta E^*_{SO}(k)$.

This project was supported by National Research Council of Taiwan. The authors are grateful to S. Y. Liang, C. P. Yu, C. Z. Chang, K. R. Wang and S. Lourdudoss for their assistance. We are benefited from the Referee by bringing us Ref. [22].

Table I. The effective spin-splitting energy versus Al composition.

| $x$ | Si-doped@AlGaN($10^{18}$cm$^{-3}$) | mobility($10^3$cm$^2$/Vs) | $n_1$($10^{12}$cm$^{-2}$) | $\Delta E^*_{SO}$(meV) |
|---|---|---|---|---|
| 0.11[17] | undoped | 9.15 | 5.40 | 2.5 |
| 0.15[16] | undoped | 7.39 | 2.4 ~ 4.2 | 4.4 ~ 4.8 |
| 0.15[14] | 5.0 | 9.54 | 7.12 ~ 7.57 | 2.7 ~ 3.6 |
| 0.22[a] | undoped | 8.10 | 8.68 ~ 9.39 | 0 ~ 10.38 |
| 0.25[13,15] | 0.17 | 6.10 | 8.16 ~ 8.39 | 4.36 ~ 11.2 |
| 0.30[16] | undoped | 6.82 | 8.0 | 0 |
| 0.31[15] | 3.0 | 4.0 | 10.30 | 5.91 |

[a] This work.

Figure captions

Fig. 1.  (a) Schematic diagram of spin-degenerate sandglass cone for a bulk wurtzite structure, which obeys the degenerate equation, $[\alpha_{wz} - \gamma_{wz}(bk_z^2 - k_{//}^2)] = 0$. (b) The temperature dependent SdH measurements before illumination, (c) the effective mass versus magnetic field and (d) the FFT spectra.

Fig. 2.  (a) The SdH measurements for different illumination times at T = 0.38 K, and (b) their FFT spectra.  (c) The temperature-dependent SdH measurements after the 12671s illumination, and (d) their FFT spectra.

Fig. 3.  (a) The SdH measurements against the illumination time at T = 0.38 K, and (b) their FFT spectra.

Fig. 4.  (a) The carrier concentrations of spin-up ($n_\uparrow$) and spin-down ($n_\downarrow$) first subbands versus total carrier concentration of first subband ($n_1$).  The effective spin splitting energy ($\Delta E^*_{SO}$) versus Fermi wave vector is shown as well.  (b) The calculated electric field ($E$) and average distance $<z>$ against total sheet carrier concentration ($n_T$).

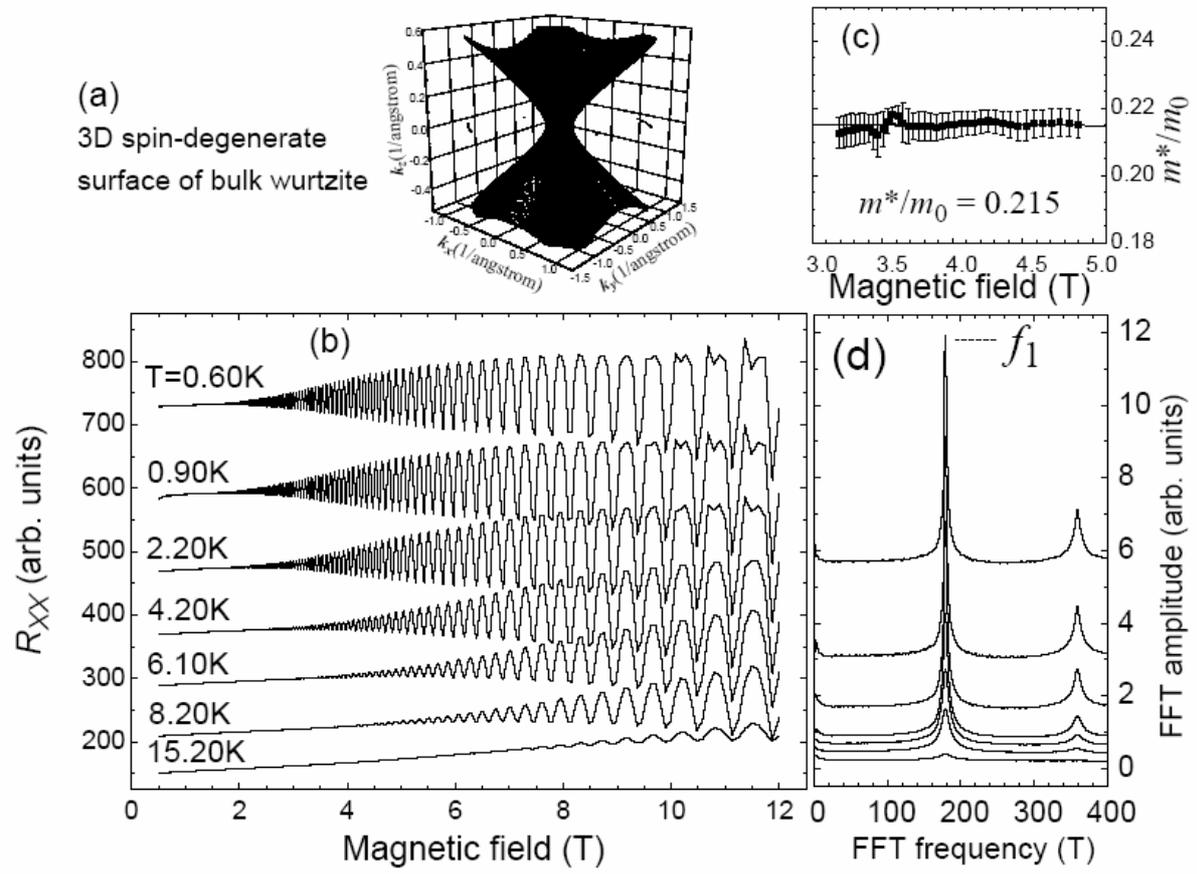

Fig. 1

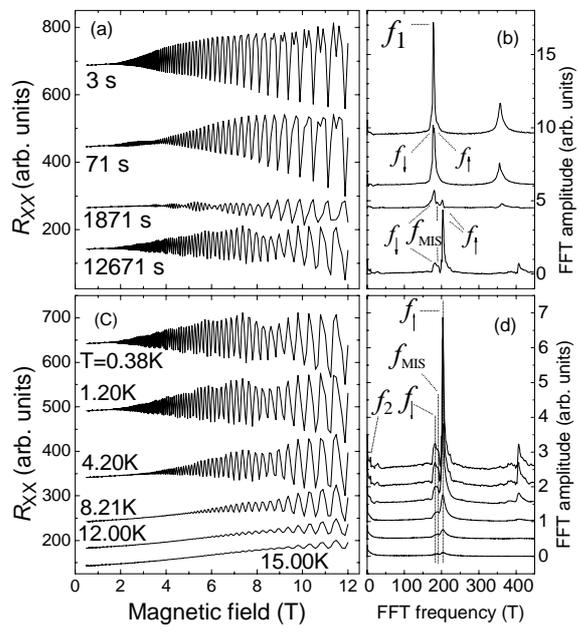

Fig. 2

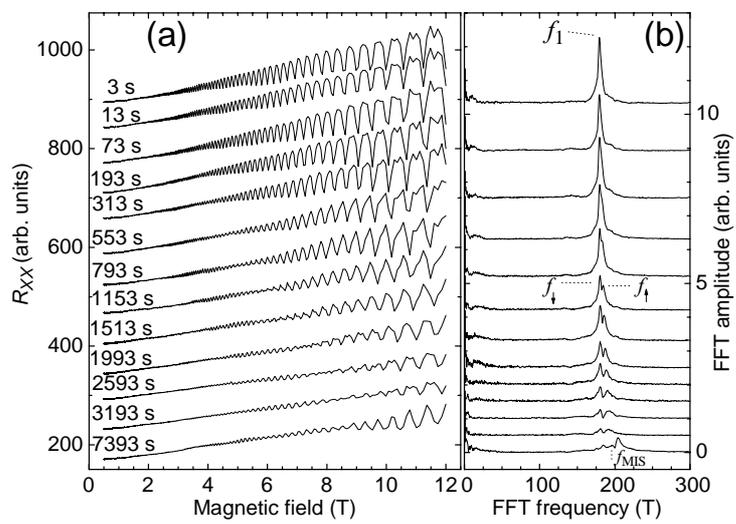

Fig. 3

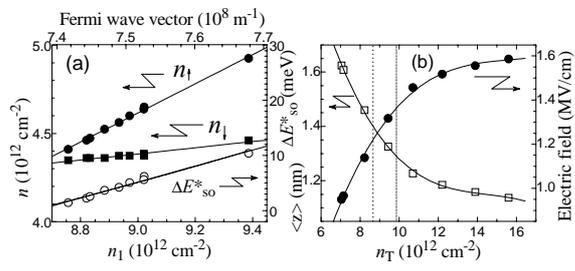

Fig. 4